\documentclass[11pt]{article}
\usepackage{moriond,epsfig}
\usepackage{wrapfig}

\def\be{\begin{equation}}
\def\ee{\end{equation}}
\def\bea{\begin{eqnarray}}
\def\eea{\end{eqnarray}}

\def\Dzkpi{D^0\to K^-\pi^+}
\def\Dzkpipiz{D^0\to K^-\pi^+\pi^0}
\def\Dzkpipipi{D^0\to K^-\pi^+\pi^+\pi^-}
\def\Dpkpipi{D^+\to K^-\pi^+\pi^+}
\def\Dpkpipipiz{D^+\to K^-\pi^+\pi^+\pi^0}
\def\Dpkspi{D^+\to K^0_S\pi^+}
\def\Dpkspipiz{D^+\to K^0_S\pi^+\pi^0}
\def\Dpkspipipi{D^+\to K^0_S\pi^+\pi^+\pi^-}
\def\Dpkkpi{D^+\to K^+K^-\pi^+}

\def\NDzDzbar{N_{D^0\bar D^0}}
\def\NDpDm{N_{D^+D^-}}

\begin{document}
\vspace*{4cm}
\title{Open Charm Physics at CLEO-c}

\author{ A. Ryd \\
(Representing the CLEO Collaboration) }

\address{Cornell University, Newman Laboratory, Ithaca NY 14853, USA}

\maketitle\abstracts{
Recent CLEO-c results on open charm physics at the $\psi(3770)$
are presented. Measurements of  
hadronic and semileptonic branching fractions 
of the $D^0$ and $D^+$ mesons are discussed as well
as the leptonic decay $D^+\to \mu^+\nu_{\mu}$ 
and determination of the $D$ meson decay constant.
}

\section{Introduction}

The CLEO-c experiment at the Cornell Electron Positron Storage Ring
has recorded 281 pb$^{-1}$ at the $\psi(3770)$. This sample
provides a very clean environment for studying decays of 
$D$ mesons. The $\psi(3770)$
produced in the $e^+e^-$ annihilation decays to pairs of $D$ mesons,
either $D^+D^-$ or $D^0\bar D^0$. In particular, the
produced $D$ mesons can not be accompanied
by any additional pions.

The analyses presented here all have in common that they use a 
tagging technique. This technique was used by the MARK III
collaboration~\cite{markiii}. In these
analyses one of the $D$ mesons is fully reconstructed in a 
hadronic final state. From energy and momentum conservation we
can predict the four-momentum of the other $D$ 
meson in the event.

We report here on measurements of the absolute
hadronic branching fractions of $D^0$ and $D^+$ mesons,
measurements of semileptonic branching fractions and the measurement
of the branching fraction for the leptonic decay 
$D^+\to\mu^+\nu_{\mu}$ and the determination of the $D$ meson decay
constant $f_D$.

\section{Absolute $D$ hadronic branching fractions}

Determination of the absolute hadronic branching fractions 
for $D$ mesons is important as they provide the normalization
for practically all $B$ decays, and as such impact for
example the determination of $|V_{cb}|$ using exclusive 
$B\to D^*\ell\nu$. The branching fractions for $D^0$
decays have been measured to about 3\% prior to CLEO-c
while $D^+$ meson branching fractions were only known
to about 6\%. The results presented here on 56 pb$^{-1}$ 
represent significant improvements to the $D^+$ 
branching fractions~\cite{dhadprl}.

This analysis makes use of a 'double tag' technique
in which we determine the number of single tags,
separately for $D$ and $\bar D$ decays,
$N_i=\epsilon_i{\cal B}_i N_{D\bar D}$ and 
${\bar N}_j=\bar \epsilon_j{\cal B}_j N_{D\bar D}$
where $\epsilon_i$ and ${\cal B}_i$ are the efficiencies
and branching fractions for mode $i$. Similarly
we reconstruct double tags where both $D$ mesons are
found. The number of double tags found is given
by $N_{ij}=\epsilon_{ij}{\cal B}_i{\cal B}_j N_{D\bar D}$
where $i$ and $j$ label the $D$ and $\bar D$ mode used
to reconstruct the event and $\epsilon_{ij}$ is the
efficency for reconstructing the final state.
Combining the two equations above we can solve for $N_{D\bar D}$
as 
$$
N_{D\bar D}={{N_i}{\bar N_j}\over N_{ij}}{\epsilon_{ij}\over \epsilon_i\bar\epsilon_j}.
$$

\begin{figure}[tb]
\begin{center}
\begin{minipage}{0.9\linewidth }
\begin{minipage}{0.72\linewidth }
\includegraphics[width=\linewidth]{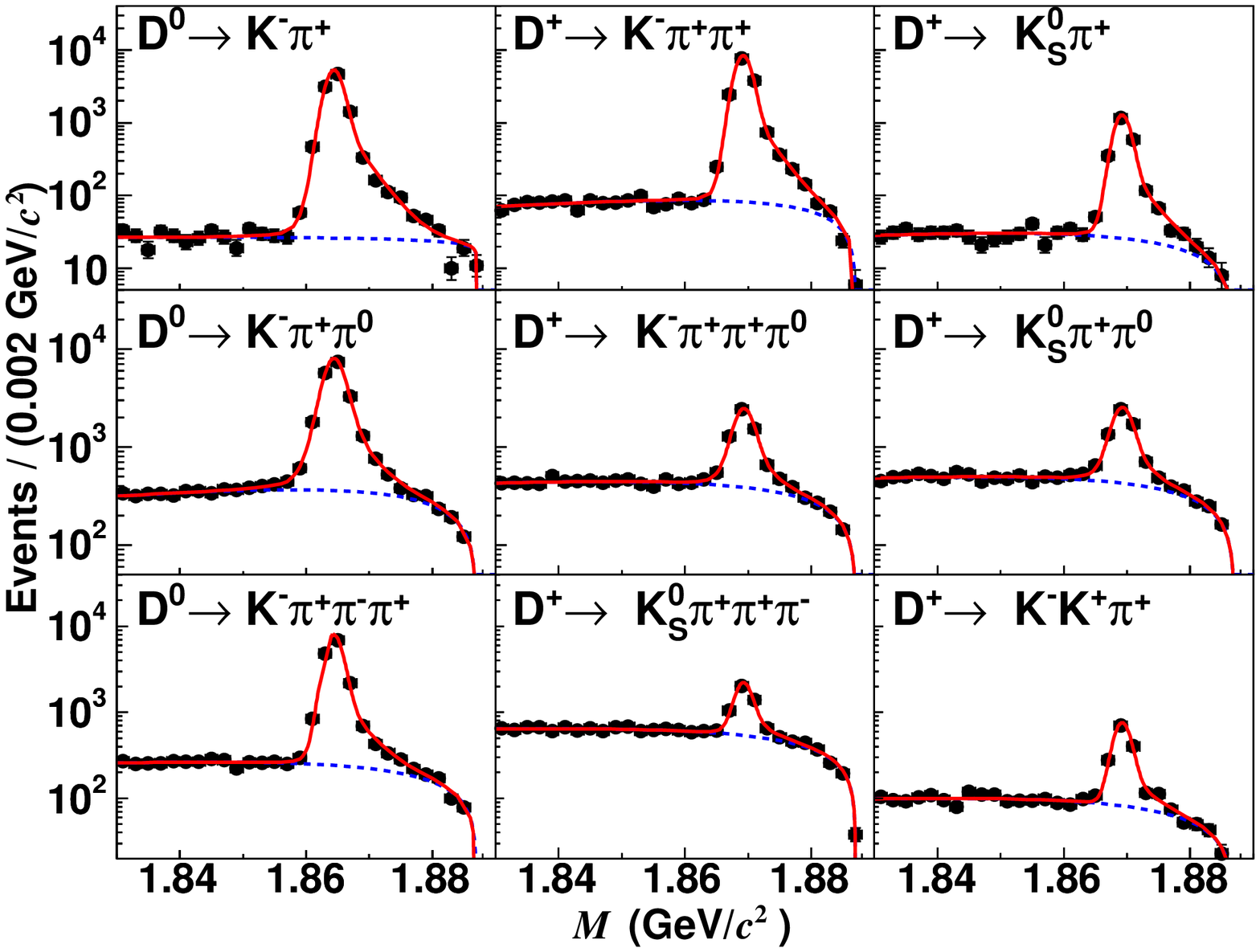}
\end{minipage}
\begin{minipage}{0.27\linewidth}
\includegraphics[width=\linewidth]{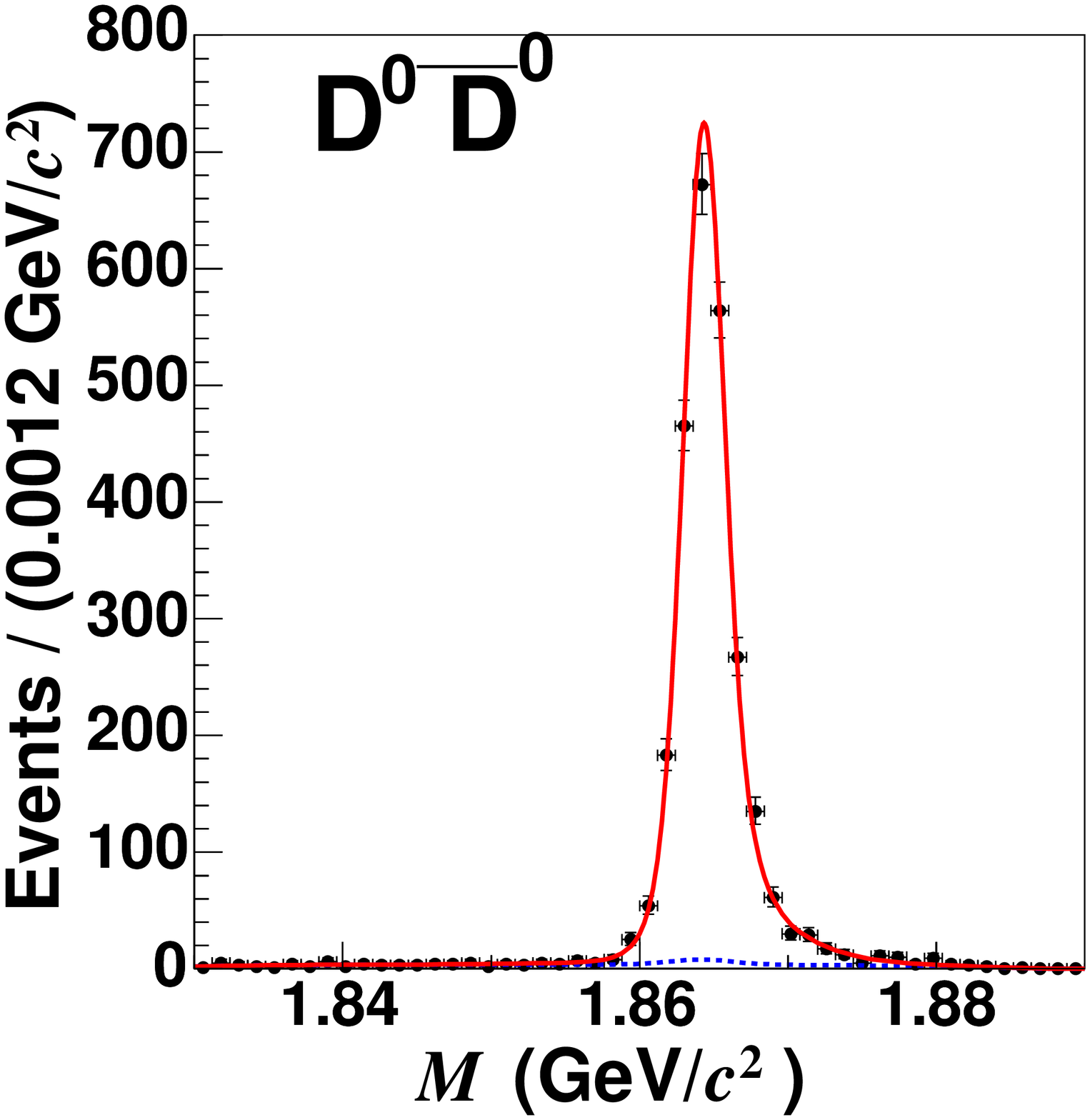}
\includegraphics[width=\linewidth]{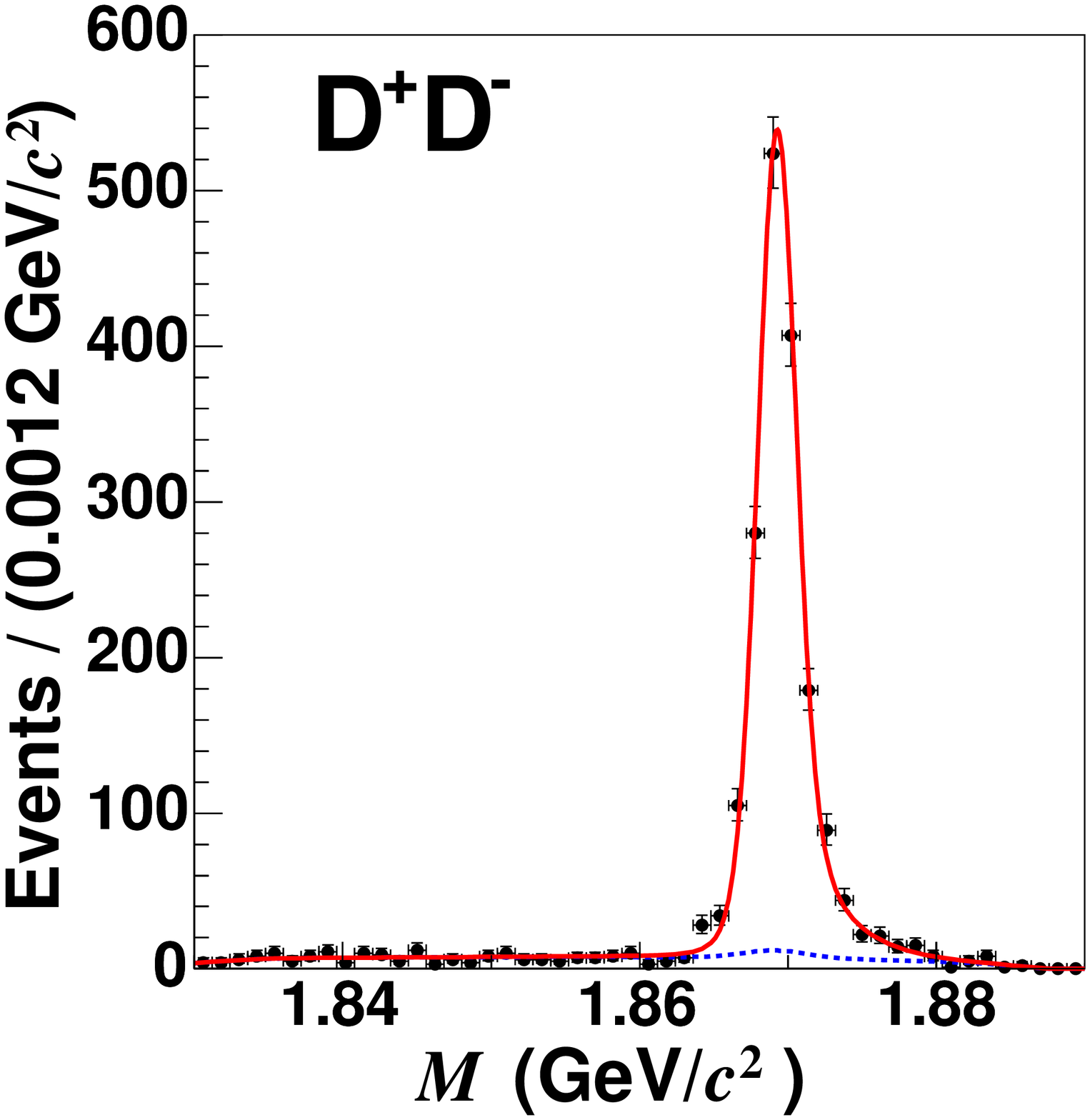}
\end{minipage}
\end{minipage}
\caption{Yields for single tags, left, and double tag yields combined
an all neutral and charged modes respectively in the right two plots. }
\label{fig:dhad_yields}
\end{center}
\end{figure}

In this analysis we make use of three $D^0$ decays 
and six $D^+$ modes
as shown in Fig.~\ref{fig:dhad_yields}.
We have a 
total of $2,484\pm 51$ neutral double tags and $1,650\pm42$ 
charged double tags.
The scale of the statistical error on the branching fractions
are set by the number of double tags, 
$\approx 2.0\%$ and $\approx 2.5\%$ for the neutral and charged modes
respectively.
The branching fractions obtained are summarized in Table~\ref{tab:dhadresults}.

\begin{table}[bt]
\small{
\caption{Fitted branching fractions and $D\bar D$ pair yields.  Uncertainties are statistical and systematic,
respectively. 
}
\label{tab:dhadresults}
\begin{center}
\begin{tabular}{lcc}
\hline\hline
Parameter & Fitted Value & PDG  \\
\hline
$\NDzDzbar$               & $(2.01\pm 0.04\pm 0.02)\times 10^5$ & --- \\
${\cal B}(\Dzkpi)$        & $(3.91\pm 0.08\pm 0.09)\%$        & $3.81\pm0.09\%$ \\
${\cal B}(\Dzkpipiz)$     & $(14.9\pm 0.3\pm 0.5)\%$          & $13.2\pm1.0\%$ \\
${\cal B}(\Dzkpipipi)$    & $(8.3\pm 0.2\pm 0.3)\%$           & $7.48\pm0.30\%$ \\
\hline
$\NDpDm$                  & $(1.56\pm 0.04\pm 0.01)\times 10^5$ & --- \\
${\cal B}(\Dpkpipi)$      & $(9.5\pm 0.2\pm 0.3)\%$           & $9.2\pm0.6\%$ \\
${\cal B}(\Dpkpipipiz)$   & $(6.0\pm 0.2\pm 0.2)\%$           & $6.5\pm1.1\%$ \\
${\cal B}(\Dpkspi)$       & $(1.55\pm 0.05\pm 0.06)\%$        & $1.42\pm0.09\%$ \\
${\cal B}(\Dpkspipiz)$    & $(7.2\pm 0.2\pm 0.4)\%$           & $5.4\pm1.5\%$ \\
${\cal B}(\Dpkspipipi)$   & $(3.2\pm 0.1\pm 0.2)\%$           & $3.6\pm0.5\%$ \\
${\cal B}(\Dpkkpi)$       & $(0.97\pm 0.04\pm 0.04)\%$        & $0.89\pm0.08\%$ \\
\hline\hline
\end{tabular}
\end{center}
}
\end{table}

\section{Semileptonic $D$ decays}

Semileptonic branching fractions have been
studied for several
Cabibbo favored and suppressed modes in
a sample of 56 pb$^{-1}$. 
In this analysis we reconstruct one $D$ meson and look at the 
recoil $D$ to identify the signal~\cite{d0semilep,dplussemilep}. 
The signal is identified
by requiring that one electron is found and the hadronic
final state is reconstructed. This means that all particles
except the neutrino is reconstructed. Using four-momentum
conservation we can infer the energy and momentum of the
neutrino. To extract the signal we form a quantity known 
as $U=E-P$ which is the energy minus the momentum for the
neutrino. For signal events this quantity should peak at 
zero.

\begin{figure}[tb]
\begin{center}
\begin{minipage}{0.9\linewidth }
\includegraphics[width=0.49\linewidth]{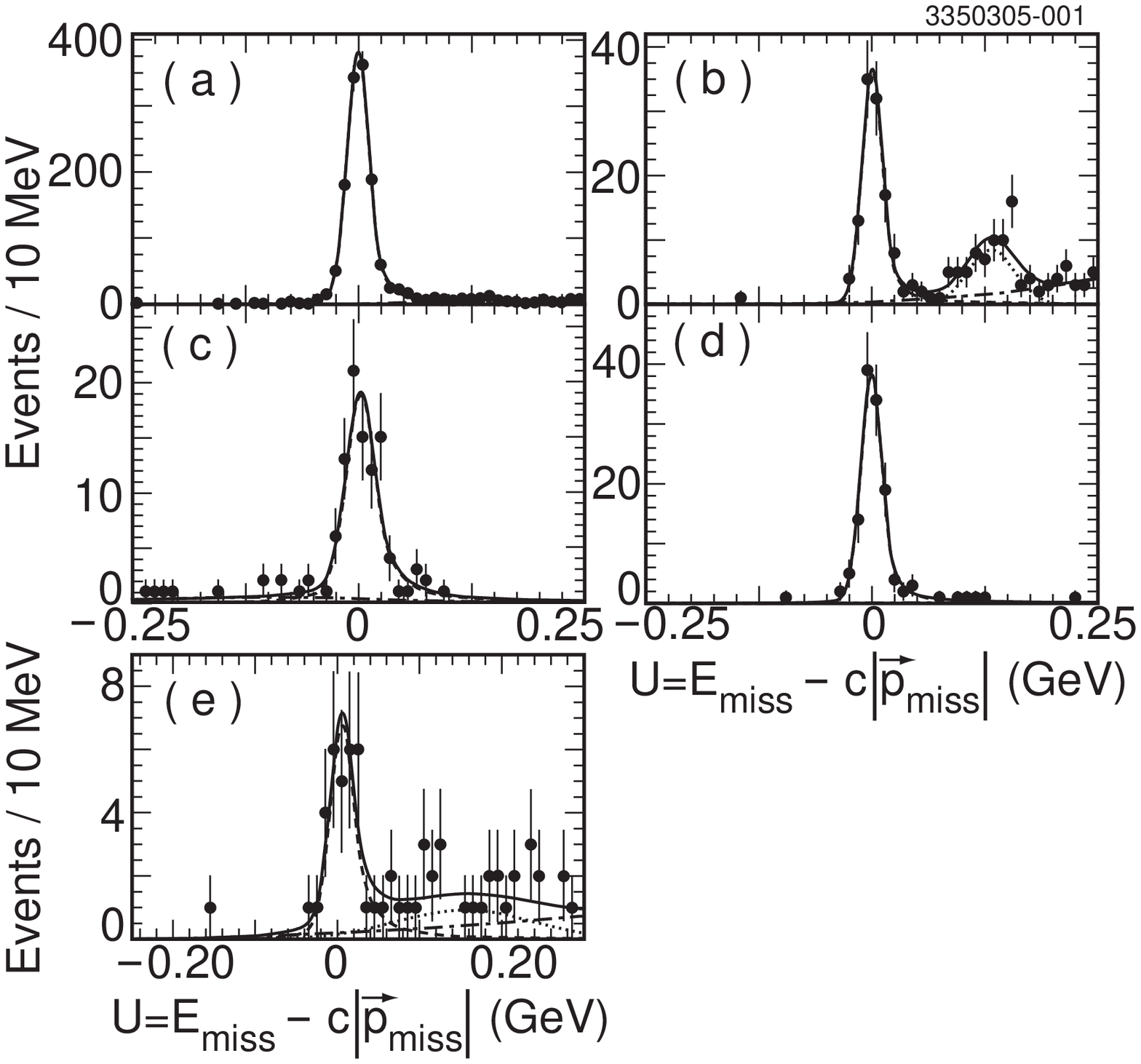}
\includegraphics[width=0.49\linewidth]{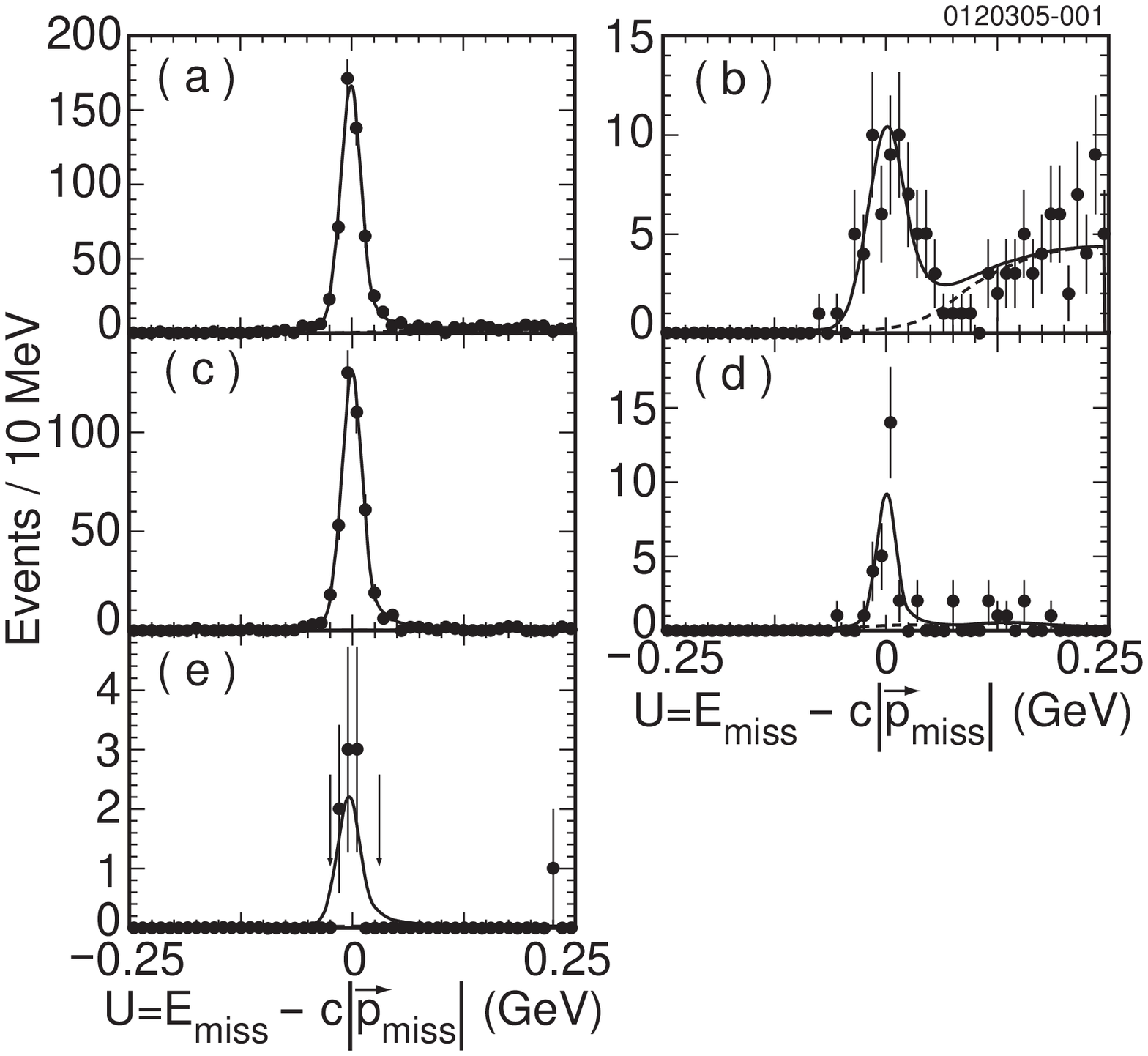}
\caption{Left: Yields of $D^0$ semileptonic decays to 
a) $K^- e^+ \nu_e$, b) $\pi^- e^+ \nu_e$, c) $K^{*-}(K^-\pi^0) e^+ \nu_e$,
d) $K^{*-}(\bar K^0\pi^-) e^+ \nu_e$, and e) $\rho^- e^+ \nu_e$.
Right: Yields of $D^+$ semileptonic decays to 
a) $\bar K^0 e^+ \nu_e$, b) $\bar K^{*0} e^+ \nu_e$, c) $\pi^0 e^+ \nu_e$,
d) $\rho^0 e^+ \nu_e$, and e) $\omega e^+ \nu_e$.
}
\label{fig:semilep}
\end{minipage}
\end{center}
\end{figure}

Figure~\ref{fig:semilep} shows the 
semileptonic yields for $D^0$ and $D^+$ decays. The 
extracted branching fractions are summarized in
Table~\ref{tab:semilepresults}.

\begin{table}[bt]
\small{
\caption{Branching fractions for semileptonic $D^0$ and $D^+$ 
meson decays.  Uncertainties are statistical and systematic,
respectively. 
}
\begin{center}
\label{tab:semilepresults}
\begin{tabular}{lcc}
\hline\hline
Mode                              & ${\cal B}(\%)$ CLEOC-c     & ${\cal B}(\%)$ PDG\\
\hline
$D^0\rightarrow \pi^- e^+ \nu_e$  & $0.26\pm0.03\pm0.01$ & $0.36\pm 0.06$  \\
$D^0\rightarrow K^- e^+ \nu_e$    & $3.44\pm0.10\pm0.10$ & $3.58\pm 0.18$  \\
$D^0\rightarrow K^{*-}(K^-\pi^0) e^+ \nu_e$ & $2.11\pm0.23\pm0.10$ & $2.15\pm 0.35$  \\
$D^0\rightarrow K^{*-}(\bar K^0\pi^-) e^+ \nu_e$ & $2.19\pm0.20\pm0.11$ & $2.15\pm 0.35$  \\
$D^0\rightarrow \rho^- e^+ \nu_e$  & $0.19\pm0.03\pm0.04$ & ---  \\
$D^+\rightarrow \pi^0 e^+ \nu_e$  & $0.44\pm0.06\pm0.03$ & $0.31\pm 0.15$  \\
$D^+\rightarrow \bar K^0 e^+ \nu_e$  & $8.71\pm0.38\pm0.37$ & $6.7\pm 0.9$  \\
$D^+\rightarrow \bar K^{*0} e^+ \nu_e$  & $5.56\pm0.27\pm0.23$ & $5.5\pm 0.7$  \\
$D^0\rightarrow \rho^0 e^+ \nu_e$  & $0.21\pm0.04\pm0.01$ & $0.25\pm0.10$  \\
$D^0\rightarrow \omega^0 e^+ \nu_e$  & $0.16^{+0.07}_{-0.06}\pm0.01$ & ---  \\
\hline\hline
\end{tabular}
\end{center}
}
\end{table}

\section{Leptonic $D^+$ decays and $f_D$}

The decay $D^+\to \mu^+\nu_{\mu}$ provides a 
direct measurement of the $D$ meson decay constant, $f_D$.
The partial width for $D^+\to \ell^+\nu_{\ell}$
is given by
$$
\Gamma(D^+\to \mu^+\nu_{\mu})={ G_{\rm F}^2 \over 8\pi} f^2_D 
       m_D m_{\ell}^2\left[ 1-{m_{\ell}^2\over m_D^2}\right]^2 |V_{cd}|^2
$$
where all factors are known except for the decay constant. 
A measurement of the branching fraction combined with the 
well known $D^+$ lifetime allows us to determine the decay 
constant. Precise measurements of the decay constant in $D$ 
and $D_s$ decays allow for precise tests of Lattice QCD.

\begin{wrapfigure}{r}{2.2in}
{\includegraphics[width=2.2in]{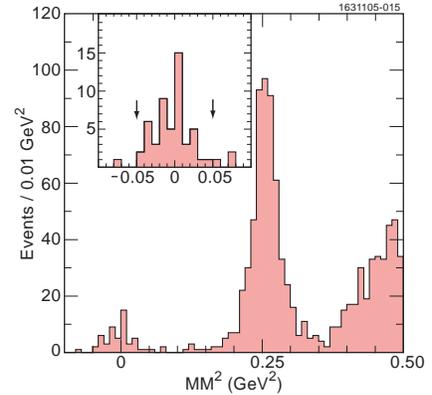}}
\caption{Missing mass squared distribution for the $D^+\to\mu^+\nu_{\mu}$ 
candidates.}
\label{fig:m2miss}
\end{wrapfigure}

This analysis reconstructs charged $D$ mesons in six modes, 
a total $158,354\pm496$ tags were reconstructed
in the 281 pb$^{-1}$ sample. We look for one, and only one additional
track from the interaction point in the event. We require this track
to be consistent with being a muon by its energy deposit 
in the electromagnetic
calorimeter (EMC), we require less than 350 MeV to be deposited. In addition,
to veto events from $D^+\to \pi^+\pi^0$, we require that there were
no extra, unmatched, showers in the EMC with an energy of over 
250 MeV.

For events that satisfy these requirements we calculate the 
missing mass square, $M^2_{\rm miss}$. This is the mass that the 
tag $D$ and muon candidate is recoiling against. For signal
events this will peak at $M^2_{\rm miss}=m^2_{\nu}=0$. 
Figure~\ref{fig:m2miss} shows the observed missing mass square 
distribution. The signal region contains 50 
events. An evaluations of the backgrounds
gives an estimate of $2.81\pm 0.30\pm 0.27$ background events in the signal 
region. Combining the signal yield of $47.2\pm 7.1^{+0.3}_{-0.8}$ events
with the number of tags and the signal detection efficiency we find 
$$
{\cal B}(D^+\to \mu^+\nu_{\mu})=(4.40\pm 0.66^{+0.09}_{-0.12})\times 10^{-4}
$$
and the decay constant $f_D=(222.6\pm16.7^{+2.8}_{-3.4})\ {\rm MeV}$.
This measurement is in good agreement with theoretical predictions. 
In particular, a recent unquenched 
lattice calculation by the Fermilab-MILC-HPQCD 
collaboration~\cite{lattice}
gives $f_D=(201\pm 3\pm 17)\ {\rm MeV}$.


\section{Summary}

The CLEO-c experiment has recorded 281 pb$^{-1}$ of $e^+e^-$
annihilation data at the $\psi(3770)$. Here results for 
hadronic branching fractions and semileptonic decays were
presented on 56 pb$^{-1}$ and the leptonic decay
was based on the full 281 pb$^{-1}$ sample. CLEO-c will
run until March 31, 2008 and we plan to record a total 
of about 750 pb$^{-1}$ at the $\psi(3770)$. We have also
recorded a sample of about 200 pb$^{-1}$  at 
$E_{\rm CM}\approx 4170$ MeV. The goal is to recored a 
sample of 750 pb$^{-1}$ at this energy for $D_s$
studies.

\section*{Acknowledgments}
This work was supported by the National Science Foundation grant
PHY-0202078  and by the Alfred P. Sloan fundation.

\end{document}